\documentclass[12pt]{article}

\RequirePackage{lineno}
\usepackage{scicite}
\usepackage{times}
\usepackage{color}
\usepackage{graphicx,epsfig,epsf}
\usepackage{graphicx}
\usepackage{amssymb}
\usepackage{rotating}
\usepackage[update,prepend]{epstopdf}
\usepackage{ulem}

\newcommand{\fs}{\mbox{$.\!\!^{\mathrm s}$}}%
\newcommand{\farcs}{\mbox{$.\!\!^{\prime\prime}$}}%
\newcommand{\degrees}{\ifmmode^{\circ}\else$^{\circ}$\fi}
\newcommand{\amin}{\ifmmode^{\prime}\else$^{\prime}$\fi}
\newcommand{\asec}{\ifmmode^{\prime\prime}\else$^{\prime\prime}$\fi}
\newcommand{\psr}{J1823$-$3021A}

\def\etal{{\it et al.}}
\def\aap{{\it Astron.~\& Astrophys.}}

\def\aj{{\it Astron.~J.}}

\def\apj{{\it Astrophys.~J.}}

\def\apjs{{\it Astrophys.~J.~Supp.}}

\def\mnras{{\it Mon.~Not.~R.~Astron.~Soc.}}
\def\nat{{\it Nature}}

\def\sci{{\it Science}}

\newenvironment{sciabstract}{%
\begin{quote} \bf}
{\end{quote}}

\newcounter{lastnote}

\title{Fermi Detection of a Luminous $\gamma$-ray Pulsar in a Globular Cluster}

\author{The Fermi LAT Collaboration$^{1,2}$\\
\normalsize{$^{1}$ All authors with their affiliations appear at the end of this
paper.}\\
\normalsize{$^{2}$ Corresponding authors: P.~C.~C.~Freire, pfreire@mpifr-bonn.mpg.de;}\\
\normalsize{T.~J.~Johnson, tyrel.j.johnson@gmail.com;D.~Parent, dmnparent@gmail.com;}\\
\normalsize{C.~Venter, Christo.Venter@nwu.ac.za.}}

\date{\today}

\begin{document} 
\linenumbers

\def\psrindex{$1.4 \pm 0.3$}
\def\psrcutoff{$1.3 \pm 0.6$}
\def\psrflux{$(1.6 \pm 0.3) \times 10^{-8}$}
\def\psreflux{$(1.1 \pm 0.1 \pm 0.2) \times 10^{-11}$}
\def\psrlumabs{$(8.4 \pm 1.6) \times 10^{34}$}
\def\psrlum{$(8.4 \pm 1.6 \pm 1.5) \times 10^{34}$}
\def\uplim{$5.5 \times 10^{-12}$}
\def\nmsp{32}

\def\fermi{{\it Fermi}}
\def\deg{$^{\circ}$}
\def\gam{$\gamma$}
\def\ergcms{erg\,cm$^{-2}$\,s$^{-1}$}
\def\ergs{erg\,s$^{-1}$}

\baselineskip24pt

\maketitle 

\begin{sciabstract}

We report the Fermi Large Area Telescope detection of
$\gamma$-ray ($>$100 megaelectronvolts) pulsations from pulsar \psr\ in the
globular cluster NGC~6624 with high significance ($\sim7\,\sigma$).
Its $\gamma$-ray luminosity $L_{\gamma}$ = \psrlumabs\, ergs per second,
is the highest observed for any millisecond pulsar (MSP) to date,
and it accounts for most of the cluster emission.
The non-detection of the cluster in the off-pulse phase
implies that its contains $<$~\nmsp\, $\gamma$-ray MSPs,
not $\sim$100 as previously estimated.
The $\gamma$-ray luminosity
indicates that the unusually large rate of change of its period
is caused by its intrinsic spin-down.
This implies that \psr\, has the largest magnetic field and is
the youngest MSP ever detected, and that such anomalous objects might be
forming at rates comparable to those of the more normal MSPs.
\end{sciabstract}


Since its launch in 2008, the Large Area Telescope (LAT) on board the
Fermi Gamma-ray Space Telescope \cite{aaa+09c} has detected
whole populations of objects previously unseen in the
$\gamma$-ray band. These include globular clusters (GCs), which are ancient
spherical groups of $\sim 10^5$ stars held together by their mutual
gravity. As a class, their $\gamma$-ray spectra show evidence for an
exponential cut-off at high energies
\cite{aaa+10c,tkh+11}, a characteristic signature of
magnetospheric pulsar emission. This is not surprising because radio
surveys have shown that GCs contain large numbers of pulsars
\cite{footnote1}, neutron stars that emit radio and
in some cases X-ray and $\gamma$-ray pulsations.

The first GC detected at $\gamma$-ray energies was
47~Tucanae \cite{aaa+09a}, soon followed by Terzan~5 \cite{khc10} 
and nine others \cite{aaa+10c,tkh+11}. Even so, no individual 
pulsars in these clusters were firmly identified in 
$\gamma$-rays\cite{footnote2}. 
GCs are more distant than most $\gamma$-ray pulsars observed in the
Galactic disk \cite{aaa+10}, thus most pulsars in them should be too
faint to be detected individually. The Fermi\ LAT lacks the spatial
resolution required to resolve the pulsars in GCs, which
tend to congregate within the inner arcminute of the cluster. 
Hence, $\gamma$-ray photons emitted by all pulsars in a given GC 
increase the photon background in the folded $\gamma$-ray profiles 
of each individual pulsar in that cluster.


One of the GCs detected at $\gamma$-ray energies is 
NGC~6624 \cite{tkh+11}, located at a distance $d = 8.4 \pm 0.6$\,kpc
from Earth 
\cite{vfo07}. With a radio flux density at 400 MHz
of $S_{400} = 16$ mJy \cite{bbl+94}, \psr\, is the brightest of the six pulsars
known in the cluster. It has been regularly timed with
the Jodrell Bank and Parkes radio telescopes since discovery, and
with the Nan\c{c}ay radio telescope since the 
launch of the Fermi\ satellite. The resulting radio ephemeris
(Table S1) describes the  measured pulse times of arrival very well
for the whole length of the Fermi\ mission, the root mean square of
the timing residuals being 0.1\% of the pulsar rotational period.

Thus we can confidently use it to assign a pulsar
spin phase $\phi$ to every $\gamma$-ray ($>$0.1\,GeV) photon arriving at the
Fermi-LAT from the direction (within 0.8$^{\circ}$) of the
pulsar. We selected photons that occurred between 4 August 2008 and
4 October 2010 that pass the ``Pass 6 diffuse''
$\gamma$-ray selection cuts \cite{aaa+09c}.
The resulting pulsed $\gamma$-ray signal (above 0.1\,GeV, Fig.~1)
is very robust, with an H-test value of 64 \cite{db10} corresponding to
6.8\,$\sigma$ significance. The data are well modeled
by a power law with spectral index \psrindex\ and an exponential
cutoff at an energy of \psrcutoff\,GeV, typical of the values found
for other $\gamma$-ray pulsars [see supporting online material (SOM)].
The two peaks are aligned, within uncertainties, with the two main
radio components at spin phases $\phi_1 = 0.01 \pm 0.01$ and
$\phi_2 = 0.64 \pm 0.01$ (Fig.~1).

The pulsed flux above
0.1\,GeV, averaged over time, is $F_{\gamma}\,=\,$\psreflux\,\ergcms,
where the first errors are statistical and the second are systematic
(SOM). The large distance of NGC~6624 implies that
\psr\ is one of the most distant $\gamma$-ray pulsars detected \cite{aaa+10}. 
This makes it the most luminous $\gamma$-ray MSP
to date \cite{LAT_MSPs}: Its total emitted power is $L_{\gamma} = 4 \pi
d^2 f_{\Omega} F_{\gamma}\,=\,$\psrlum\,$(f_{\Omega}/0.9)\,$\ergs.
We obtained the statistical uncertainty by adding the uncertainties
of $d$ and $F_{\gamma}$ in quadrature. The term $f_{\Omega}$ is
the power per unit surface across the whole sky divided by power per
unit surface received at Earth's location; detailed modeling
of the $\gamma$ and radio light curves provides a best fit centered at
0.9, but with a possible range from 0.3 to 1.8 (SOM).  

The LAT image of the region around NGC~6624 during the on-pulse 
interval ($0.60<\phi<0.67$ and $0.90<\phi<1.07$) shows a bright 
and isolated $\gamma$-ray source that is consistent with the location 
of \psr\ (Fig.~\ref{fig:NGC6624_image});
in the off-pulse region ($0.07<\phi<0.60$ and $0.67<\phi<0.90$) 
no point sources in the energy band 0.1 - 100\,GeV are
detectable.
Assuming a typical pulsar spectrum with a spectral index of 1.5 and a
cut-off energy of 3\,GeV, we derived, after scaling to the full pulse
phase, a 95\% confidence level upper limit on the point source energy
flux of \uplim\,\ergcms. Thus, \psr\ dominates
the total $\gamma$-ray emission of the cluster. The combined emission 
of all other MSPs in the cluster plus any off-pulse emission from \psr\, is
not detectable with present sensitivity. No other pulsars are detected
in a pulsation search either.

Under the assumption that the $\gamma$-ray emission originates from NGC 6624, 
\cite{tkh+11} estimated the total number of MSPs to be
$N_{\rm MSP} = 103^{+104}_{-46}$. Assuming an average $\gamma$-ray
luminosity for each MSP \cite{aaa+09a,aaa+10c}, similar to the
approximation made by \cite{tkh+11}, our off-pulse flux upper limit
implies that $N_{\rm MSP} < \nmsp$. This is consistent with the
estimate $N_{\rm MSP} = 30 \pm 15$  derived from the correlation
between $\gamma$-ray luminosity and encounter rate
\cite{aaa+10c}. Clearly, the MSP number estimate of \cite{tkh+11}
is skewed by the presence of a single bright pulsar contributing
disproportionately to its emission \cite{fg00}.
The off-pulse emission limits can also be used to constrain alternative
models for the $\gamma$-ray emission from globular
clusters, like those invoking inverse Compton (IC) radiation \cite{bs07,ccd+10}.

The spin period of \psr, 5.44 ms, is typical of MSPs.
However, its rate of change
$\dot{P}_\mathrm{obs} = +3.38 \times 10^{-18}$\,s\,s$^{-1}$ is one to two
orders of magnitude larger than for other MSPs with the exception of
J1824$-$2452A, a pulsar in the GC M28 \cite{lbm+87} that has a
similarly large $\dot{P}_\mathrm{obs}$ \cite{fbtg88}. A possible explanation
is that $\dot{P}_\mathrm{obs}$ is due mostly to the changing Doppler shift
caused by the pulsar's acceleration in the gravitational field of the
cluster along the line of sight ($a_l$):
\begin{equation}\label{eq:period}
\label{eq:acc}
\left( \frac{\dot{P}_{\rm obs}}{P} \right) =
\left( \frac{\dot{P}}{P} \right) + \frac{a_l}{c}.
\end{equation}
If the globular cluster has a reliable mass model, we could use it to
estimate lower and upper limits for $a_l$ and estimate upper and lower
limits for $\dot{P}$ \cite{fck+03}. For 
NGC~6624 the collapsed nature of its core precludes the 
derivation of a reliable mass model. Furthermore, radio 
timing (Table S1) shows that \psr\ is only $0\farcs4 \pm 0\farcs1$ (a
projected distance of $0.018\pm0.004$ pc) from the center of the
cluster \cite{gra+10}, where the values of $a_l$ can be 
largest. For this reason, it has been suggested \cite{bbl+94}
that \psr\, is a ``normal'' MSP (i.e., with small $\dot{P}$); its
large $\dot{P}_\mathrm{obs}$ being due to its acceleration
in the cluster. This conclusion was apparently strengthened by the
detection of a second derivative of the spin period
$\ddot{P} = -1.7 \times 10^{-29}$\,s$^{-1}$ \cite{hlk+04}.
This could originate in a time variation
of $a_l$ resulting from interaction with a nearby object \cite{phi92}.
If sustained it would reverse the sign of
$\dot{P}_{\rm obs}$ in $\sim$\,6000 years; suggesting again
that the large $\dot{P}_\mathrm{obs}$ is not only due to
dynamical effects, but is possibly a transient feature.

However, the total observed $\gamma$-ray emission $L_{\gamma}$ 
must represent a fraction $\eta < 1$ of the available 
rotational energy loss, $\dot{E} = 4 \pi^2 I \dot{P}/ P^3$, where
$I$ is the pulsar's moment of inertia. Although $I$ depends on
the unknown mass of the pulsar and the unknown equation of state
for dense matter, the standard asumption $I = 10^{45} \rm \, g\, cm^{2}$
is a reasonable value for a 1.4-$M_{\odot}$ (mass of the Sun)
neutron star. This implies
$\dot{P} > 3.4 \times 10^{-19}\,(f_{\Omega}/0.9)(I/10^{45}\rm g\, cm^2)^{-1}$\,s\,s$^{-1}$. Thus
even an unrealistic $\gamma$-ray efficiency $\eta = 1$ would imply
that $\dot{P}$ is already $\sim$10\% of $\dot{P}_{\rm obs}$.
If we assume instead $\dot{P} \simeq \dot{P}_\mathrm{obs}$, then
$\dot{E} = 8.3\,\times\,10^{35}$\,\ergs\ and $\eta = 0.1 \times
(f_{\Omega}/0.9)(I/10^{45}\rm g\,cm^2)^{-1}$.
Comparison with the observed $\gamma$-ray
efficiencies of other MSPs \cite{LAT_MSPs,aaa+10} shows this
to be a more reasonable range of values; $\eta \sim 0.1$ also
represents the upper limit derived for the average efficiency of
MSPs in 47 Tucanae \cite{aaa+09a}. Therefore, our $\gamma$-ray
detection of \psr\, indicates that it is unusually energetic and
that most of $\dot{P}_\mathrm{obs}$ is due to
its intrinsic spin-down.
The pulsar has other features that suggest it is indeed unusually energetic:
Its alignment of radio and $\gamma$-ray profiles has previously
only been observed for the Crab pulsar \cite{aaa+10a} and three fast,
energetic MSPs: J1939+2134 (the first MSP to be discovered),
J1959+2048 \cite{gcj+11} and J0034$-$0534 \cite{aaa+10b}. Like some of
these energetic pulsars and PSR~J1824$-$2452A, \psr\, emits giant radio
pulses \cite{kni07} and has a high 400 MHz radio luminosity of
$L_{400}\,\simeq\, 1.1$ Jy kpc$^2$ \cite{bbl+94}, the third highest among
known MSPs. However the correlation between $\dot{E}$ and radio
luminosity is far from perfect given the uncertainties in the distance
estimates, moment of inertia, beaming effects and
possibly intrinsic variations of the emission efficiencies.
Finally, J1939+2134 also has a large $\ddot{P}$ \cite{cbl+95},
which is thought to be caused by timing noise (TN),
which scales roughly with $P^{-1.1}\dot{P}$
\cite{sc10}. In the case of \psr, if $\dot{P} \simeq \dot{P}_\mathrm{obs}$,
then TN should be one order of magnitude larger than for J1939+2134;
instead its $\ddot{P}$ is $\sim 1.5 \times 10^2$
larger than that of J1939+2134. This is possible given the observed
scatter around the TN scaling law. Thus
TN might account for the $\ddot{P}$ of \psr, but this is far more
likely if $\dot{P} \simeq \dot{P}_\mathrm{obs}$.

If $\dot{P} \simeq \dot{P}_\mathrm{obs}$,
we can estimate the strength of its surface dipole magnetic
field: $B_0 = 3.2 \times 10^{19} {\rm G} \sqrt{\dot{P} P (I / 10^{45}
{\rm\,g\,cm^2})} (R/ 10{\rm\,km})^{-3} \simeq 4.3 \times
10^9$ G\cite{lk04} [where $R$ is the neutron star (NS) radius, generally
assumed to be 10 km]. MSPs are thought to start
as normal NSs with $B_0 \sim 10^{11-13}$\,G which are
then spun up by the accretion of matter and angular momentum
from a companion star. This process is thought to decrease their magnetic
field to $B_0 \sim 10^{7-9}$\,G; but the exact mechanism responsible
for this is currently not well understood. Our value of
$B_0$ shows that for \psr\, this decrease 
was not as pronounced as for other MSPs.

As accretion spins up the NS, it eventually reaches an equilibrium
spin period \cite{acrs82} given by:
\begin{equation}
\label{eq:pmin}
P_{\rm init} = 2.4 {\rm ms} \left( \frac{B_0}{10^9 {\rm G}}  \right)^{6/7}
                         \left( \frac{M}{M_{\odot}} \right)^{-5/7}
                         \left( \frac{R}{10^4 \rm m} \right)^{18/7}
                         \left( \frac{\dot{M}}{\dot{M}_{\rm Edd}}  \right)^{-3/7},
\end{equation}
where $M$ is the NS mass, $\dot{M}$ is the accretion rate
and $\dot{M}_{\rm Edd}$ is the maximum possible stable accretion rate
for a spherical configuration (known as the Eddington rate). Beyond
this, the pressure of accretion-related radiation starts preventing
further accretion. After accretion ceases, the newly formed radio MSP
will have $P_{\rm init}$ as its initial spin period.
Assuming $\dot{M} = \dot{M}_{\rm Edd}$,
$M = 1.4\,M_{\odot}$ and $R =$10~km (as in our estimates of
$B_0$), we obtain $P_{\rm init} = 1.9\,{\rm ms} ( B_0 / 10^9 {\rm
  G})^{6/7}$. For the value of $B_0$ calculated above, we get $P_{\rm
  init} = 6.6\,$ms; that is, even if accretion had proceeded at
  the Eddington rate, the pulsar would not have been
spun up to its present spin frequency. This is also the case for
the other such ``anomalous'' MSP,
J1824$-$2452A \cite{fbtg88}; for all others we have $P > P_{\rm init}$.
A possible explanation is that for these two objects $M$ and $I$
do not correspond to the assumptions above. If, for example, $\eta\,=\,0.15$,
$M\,=\,1.8\,M_{\odot}$ and
$I\,=\,1.8 \times 10^{45} \rm g\,cm^2$ \cite{wkl08} we obtain
$B_0 = 3.6 \times 10^9$ G and 
$P_{\rm init}\,=\,4.7\,{\rm ms}$. A second possibility, suggested by
eq.~\ref{eq:pmin}, is super-Eddington accretion (more precisely,
$\dot{M} > 1.6\,\dot{M}_{\rm Edd}$); this can happen for non-spherical mass
accretion. A third possibility is that the value of $B_0$ was smaller
during accretion (resulting in a smaller $P_{\rm init}$), and that 
$B_0$ has increased since then. This has been observed for some normal pulsars
\cite{elk+11}; however there is no evidence of such behavior for any
other MSPs.

In any case, the conclusion that
$\dot{P} \simeq \dot{P}_\mathrm{obs}$ implies a characteristic age
$\tau_c = P / (2 \dot{P}) = 25$ million years. This is likely an
over-estimate of the true age of the pulsar, particularly given that
$P_{\rm init}$ is likely to be similar to $P$. Thus \psr\ is likely to
be the youngest MSP ever detected; only J1824$-$2452A might have a
comparable age. Because of their large
$\dot{P}$s both objects will be observable as MSPs for a time that 
is $\sim 10^2$ shorter than the $\sim 100$ ``normal'' radio-bright
MSPs known in GCs. Statistically, this suggests that, at least in
GCs, anomalous high $B$-field
MSPs like \psr\, and J1824$-$2452A are forming at rates
comparable to those of the more ``normal'', radio-bright MSPs.



The Fermi-LAT Collaboration acknowledges support from a number 
of agencies and institutes for both development and the operation of 
the LAT as well as scientific data analysis. These include NASA and the
Department of Energy 
in the United States; Comissariat \`a l'Energie Atomique et
aux \'Energies Alternatives / Institut de Recherche sur les Lois
Fundamentales de l'Universe (CEA/IRFU) and
Institut National de Physique Nucl\'eaire et de Physique des
Particules, Centre National de la Recherche Scientifique
(IN2P3/CNRS) in France; Agenzia Spaziale Italiana (ASI) and
Instituto Nazionale di Fisica Nucleare (INFN) 
in Italy; Ministry of Education, Culture, Sports, Science and
Technology (MEXT), Energy Accelerator Research Organization (KEK),
and Japan Aerospace Exploration Agency (JAXA) in Japan;
and the K.~A.~Wallenberg 
Foundation, Swedish Research Council and the National Space Board 
in Sweden. Additional support from
Instituto Nazionale di Astrofisica (INAF) in Italy and
Centre National d'Etudes Spaciales (CNES) in France for 
science analysis during the operations phase is also gratefully acknowledged.
The Nan\c cay Radio Observatory is operated by the Paris Observatory, 
associated with the CNRS.
The Lovell Telescope is owned and operated by the University 
of Manchester as part of the Jodrell Bank Centre for Astrophysics with 
support from the Science and Technology Facilities Council of the 
United Kingdom.
Fermi LAT data, $\gamma$-ray diffuse models, and radio pulsar
ephemeris are available from the Fermi Science Support Center
(http://fermi.gsfc.nasa.gov/ssc/data/access).
We also thank the anonymous referees
and the Science editors for the very constructive suggestions.

{\bf The Fermi LAT Collaboration:}
P.~C.~C.~Freire$^{1}$, A.~A.~Abdo$^{2}$, M.~Ajello$^{3}$,
A.~Allafort$^{3}$, J.~Ballet$^{4}$,
G.~Barbiellini$^{5,6}$, D.~Bastieri$^{7,8}$, K.~Bechtol$^{3}$,
R.~Bellazzini$^{9}$, R.~D.~Blandford$^{3}$
% ...
\\

$^{1}$ Max-Planck-Institut f\"ur Radioastronomie, Auf dem H\"ugel 69, 53121 Bonn, Germany. $^{2}$ Center for Earth Observing and Space Research, College of Science, George Mason University, Fairfax, VA 22030, resident at Naval Research Laboratory, Washington, DC 20375, USA. $^{3}$ W. W. Hansen Experimental Physics Laboratory, Kavli Institute for Particle Astrophysics and Cosmology, Department of Physics and SLAC National Accelerator Laboratory, Stanford University, Stanford, CA 94305, USA. $^{4}$ Laboratoire AIM, CEA-IRFU/CNRS/Universit\'e Paris Diderot, Service d'Astrophysique, CEA Saclay, 91191 Gif sur Yvette, France. $^{5}$ Istituto Nazionale di Fisica Nucleare, Sezione di Trieste, I-34127 Trieste, Italy. $^{6}$ Dipartimento di Fisica, Universit\`a di Trieste, I-34127 Trieste, Italy. $^{7}$ Istituto Nazionale di Fisica Nucleare, Sezione di Padova, I-35131 Padova, Italy. $^{8}$ Dipartimento di Fisica ``G. Galilei", Universit\`a di Padova, I-35131 Padova, Italy. $^{9}$ Istituto Nazionale di Fisica Nucleare, Sezione di Pisa, I-56127 Pisa, Italy. $^{10}$ Istituto Nazionale di Fisica Nucleare, Sezione di Perugia, I-06123 Perugia, Italy. $^{11}$ Dipartimento di Fisica, Universit\`a degli Studi di Perugia, I-06123 Perugia, Italy. $^{12}$ Dipartimento di Fisica ``M. Merlin" dell'Universit\`a e del Politecnico di Bari, I-70126 Bari, Italy. $^{13}$ Istituto Nazionale di Fisica Nucleare, Sezione di Bari, 70126 Bari, Italy. $^{14}$ Laboratoire Leprince-Ringuet, \'Ecole polytechnique, CNRS/IN2P3, Palaiseau, France. $^{15}$ Institut de Ci\`encies de l'Espai (IEEE-CSIC), Campus UAB, 08193 Barcelona, Spain. $^{16}$ Columbia Astrophysics Laboratory, Columbia University, New York, NY 10027, USA. $^{17}$ INAF-Istituto di Astrofisica Spaziale e Fisica Cosmica, I-20133 Milano, Italy. $^{18}$ NASA Goddard Space Flight Center, Greenbelt, MD 20771, USA. $^{19}$ Center for Research and Exploration in Space Science and Technology (CRESST) and NASA Goddard Space Flight Center, Greenbelt, MD 20771, USA. $^{20}$ Department of Physics and Center for Space Sciences and Technology, University of Maryland Baltimore County, Baltimore, MD 21250, USA. $^{21}$ Artep Inc., 2922 Excelsior Springs Court, Ellicott City, MD 21042, resident at Naval Research Laboratory, Washington, DC 20375, USA. $^{22}$ National Research Council Research Associate, National Academy of Sciences, Washington, DC 20001, resident at Naval Research Laboratory, Washington, DC 20375, USA. $^{23}$ ASI Science Data Center, I-00044 Frascati (Roma), Italy. $^{24}$  Laboratoire de Physique et Chimie de l'Environnement, LPCE UMR 6115 CNRS, F-45071 Orl\'eans Cedex 02, and Station de radioastronomie de Nan\c{c}ay, Observatoire de Paris, CNRS/INSU, F-18330 Nan\c{c}ay, France. $^{25}$ Laboratoire Univers et Particules de Montpellier, Universit\'e Montpellier 2, CNRS/IN2P3, Montpellier, France. $^{26}$ Department of Physics and Astronomy, Sonoma State University, Rohnert Park, CA 94928-3609, USA. $^{27}$ Space Science Division, Naval Research Laboratory, Washington, DC 20375-5352, USA. $^{28}$ Santa Cruz Institute for Particle Physics, Department of Physics and Department of Astronomy and Astrophysics, University of California at Santa Cruz, Santa Cruz, CA 95064, USA. $^{29}$ Universit\'e Bordeaux 1, CNRS/IN2p3, Centre d'\'Etudes Nucl\'eaires de Bordeaux Gradignan, 33175 Gradignan, France. $^{30}$ Jodrell Bank Centre for Astrophysics, School of Physics and Astronomy, The University of Manchester, M13 9PL, UK. $^{31}$ Department of Physical Sciences, Hiroshima University, Higashi-Hiroshima, Hiroshima 739-8526, Japan. $^{32}$ Agenzia Spaziale Italiana (ASI) Science Data Center, I-00044 Frascati (Roma), Italy. $^{33}$ INAF Istituto di Radioastronomia, 40129 Bologna, Italy. $^{34}$ Max-Planck-Institut f\"ur Kernphysik, D-69029 Heidelberg, Germany. $^{35}$ Landessternwarte, Universit\"at Heidelberg, K\"onigstuhl, D 69117 Heidelberg, Germany. $^{36}$ Center for Space Plasma and Aeronomic Research (CSPAR), University of Alabama in Huntsville, Huntsville, AL 35899, USA. $^{37}$ Science Institute, University of Iceland, IS-107 Reykjavik, Iceland. $^{38}$ CSIRO Astronomy and Space Science, Australia Telescope National Facility, Epping NSW 1710, Australia. $^{39}$ College of Science, Ibaraki University, 2-1-1, Bunkyo, Mito 310-8512, Japan. $^{40}$ Research Institute for Science and Engineering, Waseda University, 3-4-1, Okubo, Shinjuku, Tokyo 169-8555, Japan. $^{41}$ CNRS, IRAP, F-31028 Toulouse cedex 4, France. $^{42}$ GAHEC, Universit\'e de Toulouse, UPS-OMP, IRAP, Toulouse, France. $^{43}$ Istituto Nazionale di Fisica Nucleare, Sezioine di Torino, I-10125 Torino, Italy. $^{44}$ Yukawa Institute for Theoretical Physics, Kyoto University, Kitashirakawa Oiwake-cho, Sakyo-ku, Kyoto 606-8502, Japan. $^{45}$ Funded by contract ERC-StG-259391 from the European Community. $^{46}$ Department of Physics and Department of Astronomy, University of Maryland, College Park, MD 20742, USA. $^{47}$ Istituto Nazionale di Fisica Nucleare, Sezione di Roma ``Tor Vergata", I-00133 Roma, Italy. $^{48}$ Department of Physics, Boise State University, Boise, ID 83725, USA. $^{49}$ Hiroshima Astrophysical Science Center, Hiroshima University, Higashi-Hiroshima, Hiroshima 739-8526, Japan. $^{50}$ Institute of Space and Astronautical Science, JAXA, 3-1-1 Yoshinodai, Chuo-ku, Sagamihara, Kanagawa 252-5210, Japan. $^{51}$ Max-Planck Institut f\"ur extraterrestrische Physik, 85748 Garching, Germany. $^{52}$ Max-Planck-Institut f\"ur Physik, D-80805 M\"unchen, Germany. $^{53}$ National Radio Astronomy Observatory (NRAO), Charlottesville, VA 22903, USA. $^{54}$ Institut f\"ur Astro- und Teilchenphysik and Institut f\"ur Theoretische Physik, Leopold-Franzens-Universit\"at Innsbruck, A-6020 Innsbruck, Austria. $^{55}$ Department of Physics, University of Washington, Seattle, WA 98195-1560, USA. $^{56}$ NYCB Real-Time Computing Inc., Lattingtown, NY 11560-1025, USA. $^{57}$ Department of Physics, Center for Cosmology and Astro-Particle Physics, The Ohio State University, Columbus, OH 43210, USA. $^{59}$ Department of Chemistry and Physics, Purdue University Calumet, Hammond, IN 46323-2094, USA. $^{59}$ Argelander-Institut f\"ur Astronomie, Universit\"at Bonn, 53121 Bonn, Germany. $^{60}$ Department of Physics, Willamette University, Salem, OR 97031, USA. $^{61}$ Instituci\'o Catalana de Recerca i Estudis Avan\c{c}ats (ICREA), Barcelona, Spain. $^{62}$ NASA Postdoctoral Program Fellow, USA. $^{63}$ Centre for Space Research, North-West University, Potchefstroom Campus, Private Bag X6001, 2520 Potchefstroom, South Africa. $^{64}$ Consorzio Interuniversitario per la Fisica Spaziale (CIFS), I-10133 Torino, Italy. $^{65}$ Dipartimento di Fisica, Universit\`a di Roma ``Tor Vergata", I-00133 Roma, Italy. $^{66}$ Department of Physics, Stockholm University, AlbaNova, SE-106 91 Stockholm, Sweden. $^{67}$ The Oskar Klein Centre for Cosmoparticle Physics, AlbaNova, SE-106 91 Stockholm, Sweden.


\clearpage

\begin{figure}[htbp]
  \begin{center}
    \includegraphics[width=12cm]{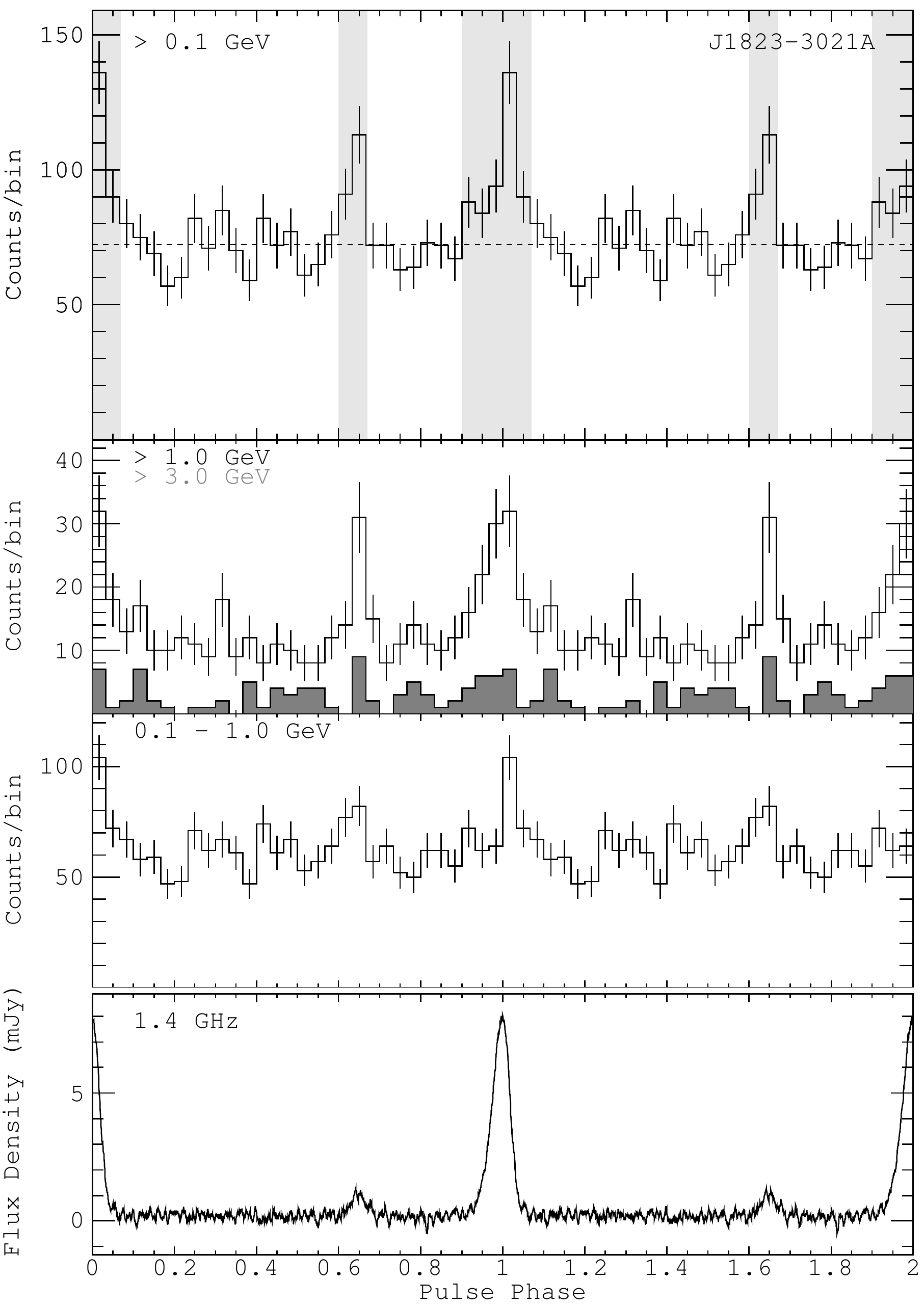}
  \end{center}
  \caption{\label{fig:NGC6624a}
  Phase-aligned radio and $\gamma$-ray profiles for \psr. (Bottom) 
  Nan\c{c}ay 1.4\,GHz radio profile. (Top and middle) 
  $\gamma$-ray profiles obtained with the Fermi-LAT in different energy 
  bands. The dark histogram is for events with $E > 3.0$GeV.
  The $\gamma$-ray background for the 0.1\,GeV light curve was 
  estimated from a surrounding ring, and it is indicated by the
  dashed horizontal line in the top panel. The highlighted area there
  shows the on-pulse region selection.
}
\end{figure}

\clearpage

\begin{figure}[htbp]
  \begin{center}
    \includegraphics[width=16cm]{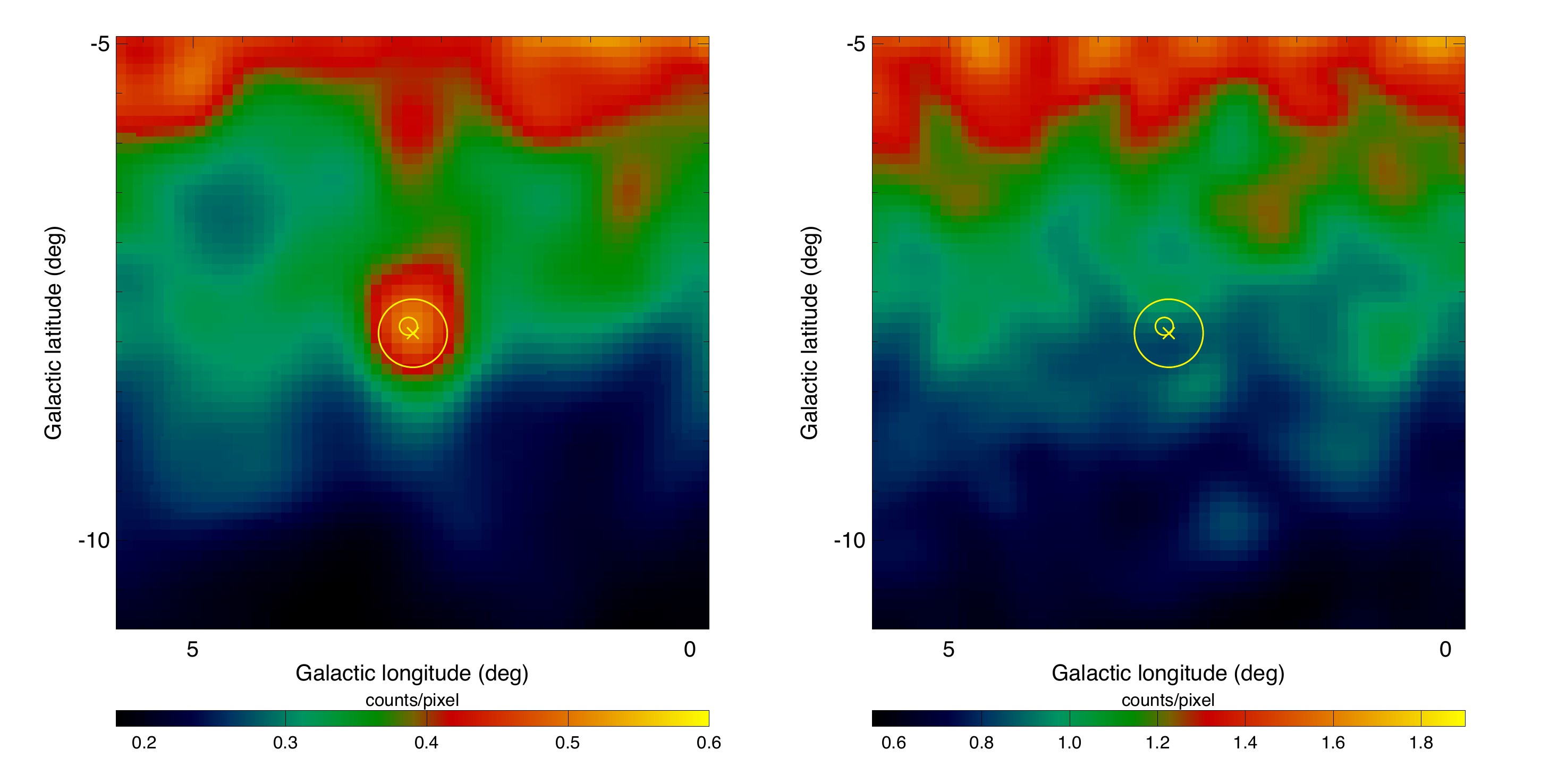}
  \end{center}
  \caption{\label{fig:NGC6624_image}Fermi\ LAT $\gamma$-ray count map 
  above 100~MeV for \psr\ during the on-pulse (Left) and off-pulse (Right)
  regions, as defined 
  in Fig. 1. The $6^{\circ}$ by $6^{\circ}$ region is centered on the pulsar 
  position (cross). The map was adaptively smoothed by imposing
  minimum signal-to-noises ratios of 13 and 16 for the on- and off-pulse regions,
  respectively. The large circle indicates the tidal radius  of NGC~6624. The small circle
  shows the 99\% confidence region for the location of the $\gamma$-ray source.}
\end{figure}


\newpage

\section*{Supporting Online Material}


\section*{Observations and data analysis}

\subsection*{Radio Timing analysis}

With the express purpose of supporting the Fermi mission \cite{sgc+08},
\psr\ is observed approximately 3 times per month with the 76-m Lovell
telescope \cite{hobbs04}, using a 64~MHz band centered at 1404~MHz
connected to an analog filterbank. Since mid-2009 observations have
also been performed using a digital filterbank backend with $1024
\times 0.5$\,MHz channels of which approximately 250\,MHz is
used. Highly precise timing measurements are also conducted with the
Nan\c{c}ay radio telescope
\cite{ctd+09}. These have included regular observations of
\psr\,since mid-2006. Approximately every two months, the pulsar is observed
for 1 hour at 1.4\,GHz. A 128\,MHz bandwidth is coherently dedispersed
using powerful GPUs (Graphics Processing Units). A total of 104 pulse times 
of arrival (TOAs) were obtained from the two telescopes between mid-2006 
and mid-2010. The
\textsc{Tempo2} timing package \cite{hem06} was used to build the
timing solution, which includes the pulsar rotation frequency and its
derivatives, the dispersion measure, and the pulsar position. The
post-fit residuals are characterized by a weighted rms of
7.3\,$\mu$s. The resulting parameters are summarized in Table~S1.
No trends are noticeable in the post-fit residuals.

\subsection*{Fermi LAT data analysis}

We have observed \psr\ with the Large Area Telescope aboard
Fermi from 2008 August 4, when the satellite began scanning-mode
operations, to the end of the validity range of the pulsar radio
ephemeris  (2010 October 14). The data analysis presented in this
paper has been performed using the LAT Science Tools package 09-21-00
and the P6 V3 Diffuse instrument response functions (IRFs). Events tagged
``Pass 6 diffuse'' having the highest probability of being \gam-ray
photons \cite{aaa+09} and coming from zenith angles $<100^\circ$ (to
reject atmospheric \gam-rays from the Earth's limb) were
used. Additionally, a rotational phase was assigned to each selected
LAT event using the radio ephemeris as an input to the Fermi
plugin \cite{rkp+11} distributed with the \textsc{Tempo2} pulsar
timing software. 

Using the {\tt pyLikelihood} likelihood fitting tool with the
NewMinuit optimizer, we performed a binned spectral analysis to
determine the energy flux and the spectral shape of the source. Events
in the range 0.1--100\,GeV were extracted from a
20\deg$\times$\,20\deg\ square region of interest (ROI) centered on
the pulsar position. To reduce the effect of the Earth's atmospheric emission,
the time intervals when the Earth was appreciably within the field of
view (specifically, when the center of the field of view was more than
52\deg\ from the zenith) were excluded from this analysis. The
Galactic diffuse emission was modeled using the \textit{gll\_iem\_v02}
map cube, while the extragalactic emission and residual instrument
backgrounds were modeled jointly by the isotropic component
\textit{isotropic\_iem\_v02}. These two models are available from the
Fermi Science Support Center\footnote{http://fermi.gsfc.nasa.gov/ssc/data/access/lat/BackgroundModels.html}. 
In addition, all the sources found in an internal catalog based on 18 
months of data (similar to \cite{LAT_1FGL}) above the background with 
significances $>5\sigma$ and within $20^\circ$
from the pulsar were included in the model. Sources were modeled with
a power law spectrum, except for pulsars for which a power law with an
exponential cut-off was used \cite{LAT_psrcat}. Sources more than
5$^\circ$ from the pulsar were assigned fixed spectra taken from the
source catalog. Spectral parameters for sources within 5$^\circ$
of the pulsar were left free for the analysis. 

The pulsar location at the core of NGC~6624 is located just outside the 
99\% statistical error contour of the $\gamma$-ray source 1FGL~J1823.4$-$3009 
\cite{LAT_1FGL}, based on an analysis of the first 11 months of the
LAT survey data. Therefore, \cite{LAT_GC}  did not establish an
association between 1FGL~J1823.4$-$3009 and the globular
cluster. \cite{TAM_GC}, using a larger dataset, showed that the
$\gamma$-ray position lies within the (20.55$\amin$) tidal radius of
the cluster. To check the association, we first reevaluated  the
position of the $\gamma$-ray source 1FGL~J1823.4$-$3009, using the
on-pulse ($0.60<\phi<0.67$ and $0.90<\phi<0.07$) segment of the pulsar
rotational phase (to improve the signal-to-noise ratio), to be
($\alpha_{2000} = 275.87$\,\deg, $\delta_{2000} = -30.29$\,\deg) with
a 99\% confidence error radius of 0.09\,\deg. This places the pulsar
radio position just inside the 99\% error contour of the $\gamma$-ray source. We also
investigated the off-pulse window and detected a $\sim 4\,\sigma$ point
source which coincides with the position of the radio source
NVSS~J182324$-$300311 ($\alpha_{2000} = 275.8515$\,\deg,
$\delta_{2000} = -30.053278$\,\deg). The signal, only observed during
the first year of the Fermi mission, is located 0.11\,\deg\ away from the position 
of 1FGL~J1823.4$-$3009 and 0.31\,\deg\ away from NGC~6624. It is likely that
1FGL~J1823.4$-$3009, located between NVSS~J182324$-$300311 and NGC~6624, 
includes contributions from both of these sources. Using the second
year of data, when the nearby faint source is off, we localize the
$\gamma$-ray source corresponding to the pulsar to ($\alpha_{2000} =
275.93$\,\deg, $\delta_{2000} = -30.34$\,\deg) with a 68\% error
counter radius of 0.07\,\deg. This position is consistent with the
radio pulsar position. We then fitted the spectrum of the pulsar at
the radio pulsar position using a power law with an exponential
cut-off. Figure~S1 shows both the fit between 0.1 and 30\,GeV (solid
lines) and the spectral points derived from likelihood fits to each
individual energy band in which it was assumed the pulsar had a
power-law spectrum.



\section*{Light Curve Modeling}

There have been two major contenders for modeling the high-energy (HE)
radiation (roughly 100 MeV to 10 GeV range) from pulsars, those which
assume that the observed $\gamma$-rays are emitted near the stellar
surface \cite{Daugherty82,DH96} above the magnetic polar cap and those
which assume the $\gamma$-rays originate primarily in the outer
magnetosphere near the light cylinder (the distance from the rotation
axis $\vec{\Omega}$ at
which the co-rotation equals the speed of light).  Both classes of
models assume that the HE $\gamma$-rays are curvature radiation from
highly-relativistic electrons/positrons in the radiation-reaction
regime.  The two most common outer-magnetospheric emission models are
the outer gap (OG; \cite{CHR86a,CHR86b,Romani96}) and the slot gap
\cite{Arons83} models.  For our purposes we took the two-pole caustic
(TPC; \cite{Dyks03}) model to be a geometric realization of the slot
gap.  TPC and OG models both assume that the emitting electrons are
accelerated up to high altitudes in narrow gaps along the last-open
field lines. The OG model only allows acceleration above the
null-charge surface (NCS; where $\vec{\Omega}\cdot\vec{B}\ =\ 0$)
whereas in the TPC model electrons are accelerated from the stellar
surface. The HE pulse profiles in these outer-magnetospheric models
are the result of the accumulation of photons in narrow phase bands
due to a combination of three effects: aberration (change of photon
direction due to the high corotation velocity), time-of-flight delays
(photons produced at higher altitudes will reach an observer earlier
than those coming from lower altitudes), as well as the magnetic field
line curvature (photons are assumed to be created tangential to the
local magnetic field line in the corotating frame, and their direction
is therefore very sensitive to the magnetic field geometry). This is
referred to as caustic emission \cite{Morini83}.

TPC and OG models are generally used in conjunction with a
low-altitude radio cone beam geometry (e.g., \cite{Atlas,VHG09}). Due
to the difference in altitude of the radio and $\gamma$-ray emission,
there will be a phase lag between the radio and $\gamma$-ray profiles.
Polar-cap (e.g., \cite{DH96}) $\gamma$-ray emission models do predict much smaller
phase lags but, due to the large open field line region of MSPs, they
cannot produce the narrow peaks observed in the $\gamma$-ray light
curve of PSR~\psr. The phase-alignment of PSR~\psr's radio and
$\gamma$-ray light curves argues for overlapping $\gamma$-ray and
radio emission regions. To reproduce the phase-aligned light curves
we used altitude-limited versions of the TPC and OG models (alTPC and
alOG, respectively) which were first introduced to model the light
curves of the MSP PSR J0034$-$0534 \cite{Abdo10-J0034}.  These are
very similar to the standard TPC and OG models, except that the
minimum and maximum radii of the radio emission region as well as the
maximum radius of the $\gamma$-ray emission region are free parameters
(the minimum $\gamma$-ray emission radius being set by the standard
models). Therefore, both radio and $\gamma$-ray photons originate in a
TPC or OG-like structure, with a significant amount of overlap between
the two emitting regions leading to phase-aligned profiles. This
implies that the radio emission is also caustic in nature, supported
by polarimetric observations which find 0\% linear polarization for
PSR~J1823$-$3021A \cite{Stairs99}. Conversely, these models provide a
framework to constrain the respective radio and $\gamma$-ray emission
geometries when comparing the model light curves to the data.

We have simulated $\gamma$-ray and radio light curves using alTPC and
alOG models with a spin period $P$ = 1.5 ms, steps of 1\deg\ in magnetic
inclination angle ($\alpha$) and viewing angle ($\zeta$), 0.05 in
accelerating emission layer width ($w$, normalized to the opening
angle of the polar cap), and 0.10 (in units of $R_{\rm LC}\ =
c P/(2\pi)$) for the emission altitudes.  The spin period used in
the simulation is less than that of PSR J1823$-$3021A (5.44 ms) but
this quantity enters the simulation mainly through the size of the
polar cap.  Using models with a shorter period will, at most,
overestimate any predicted off-pulse region. We have developed a
Markov chain Monte Carlo (MCMC) maximum likelihood technique to
jointly fit the $\gamma$-ray and radio profiles and pick the best-fit
paramters \cite{Johnson11}. We fit the $\geq$\,500\,MeV $\gamma$-ray
light curve in 60 bins and rebinned the radio profile to 60 bins. For the 
$\gamma$-ray models the minimum emission altitudes 
(R$_{min}^{\gamma}$) were specified as described previously while 
the maximum emission altitudes were allowed to be free under the 
constraint R$_{max}^{\gamma}$ $\geq$ 0.7 R$_{\rm LC}$.
The radio emission altitudes are unconstrained save
that R$_{max}^{\rm R}\ >$ R$_{min}^{\rm R}$.  The best-fit parameters
for both models are given in Table S2; the likelihood does not prefer
one model over another.  The alTPC model has best-fit gap widths of
0.0. This is unphysical and should be taken to indicate that the true
gap widths are between 0.0 and 0.05. Following \cite{VHG09} we 
can estimate the beaming factor ($f_{\Omega}$) for both models 
using Eq.~4 of \cite{Atlas}, see Table S2. Presently we are unable to 
provide reliable uncertainty estimates for our model predictions and, thus, can 
not propagate any uncertainty on $f_{\Omega}$ into the uncertainty on
$L_{\gamma}$. However, while PSR J1823$-$3021A stands out in some
respects, the shape of the  observed HE light curve is very typical of
known $\gamma$-ray MSPs. The best-fit  geometries of \cite{Johnson11}
to these MSPs yield values of $f_{\Omega}$ from approximately  0.3 to
1.8, with mean of 0.81 and rms of 0.36. We therefore expect the $f_{\Omega}$ 
value for \psr\, to be similar to the geometries which, in these models, produce 
``typical'' $\gamma$-ray light curves. The high and low tails of this
distribution suggest that the $\gamma$-ray efficiency could reasonably
be anywhere from 3 to 20\% but neither extreme affects the conclusion
of the main text that most of the observed $\dot{P}$ is intrinsic to
the pulsar.

\newpage

\begin{table*}
\begin{center}
\begin{tabular}{l l}
 \hline
 \hline
  \multicolumn{2}{c}{{\bf Timing parameters}} \\
 \hline
 \hline
  Right Ascension, $\alpha$ (J2000) \dotfill & 18$^{\rm h}\;$23$^{\rm m}\;$40$\fs$48466(4) \\
  Declination, $\delta$ (J2000) \dotfill & $-30\degrees\;$21$\amin\;$39$\farcs$988(4) \\
  Solar System Ephemeris \dotfill & DE 405 \\
  Reference time scale       \dotfill & TDB \\
  Reference time (MJD)       \dotfill & 54939\\
  Spin Frequency, $\nu$ (Hz)                  \dotfill & 183.823389814514(7) \\
  First derivative of $\nu$, $\dot{\nu}$ ($10^{-15}$ Hz s$^{-1}$) \dotfill & $-$114.1351(4) \\
  Second derivative of $\nu$, $\ddot{\nu}$ ($10^{-25}$ Hz s$^{-2}$) \dotfill & 5.8(1) \\
  Dispersion Measure, DM (cm$^{-3}$ pc)       \dotfill & 86.864(9) \\
  Validity Range (MJD) \dotfill & 53773.35 -- 55483.67 \\
  RMS Timing Residuals ($\mu$s) \dotfill & 7.3 \\
  \hline
  \hline
\end{tabular}
\label{tab:parameters}
\end{center}
{\bf Table~S1.} Timing parameters for \psr. The center of the globular cluster is
located at $\alpha\,=\,18^{\rm h} 23^{\rm m} 40\fs51 \pm 0\fs008$,
$\delta\,=\,-30\degrees 21\amin 39\farcs7 \pm 0\farcs1$.
\end{table*}

\begin{sidewaystable}
\begin{center}
\begin{tabular}{ccccccccccc}
\hline
\hline
  \multicolumn{11}{c}{{\bf Best-fit Light Curve Parameters}} \\
\hline
\hline
\\[-10pt]
Model & -log(Like)& f$_{\Omega}$ & $\alpha$ (\deg) & $\zeta$ (\deg)& $\Phi$ & $w_{\gamma}$ & $w_{\rm R}$ & R$_{max}^{\gamma}$ & R$_{min}
^{\rm R}$ & R$_{max}^{\rm R}$\\
\hline
\\[-10pt]
alTPC & 173.8 & 0.9 & 51 & 68 & 0.017 & 0.00 & 0.00 & 1.0 & 0.2 & 0.9\\
alOG & 173.6 & 0.9 & 69 & 68 & 0.017 & 0.05 & 0.10 & 1.2 & $max\lbrace \rm R_{NCS},0.2\rbrace$ & 0.8\\
\hline
\hline
\end{tabular}
\label{tab:lcfit}
\end{center}
{\bf Table~S2.} Results of MCMC maximum likelihood fits to the $\gamma$-ray and radio light curves of PSR~J1823$-$3021A using the alTPC 
and alOG models.
\end{sidewaystable} 

\newpage
\begin{figure}[htbp]
  \begin{center}
     \includegraphics[width=16cm]{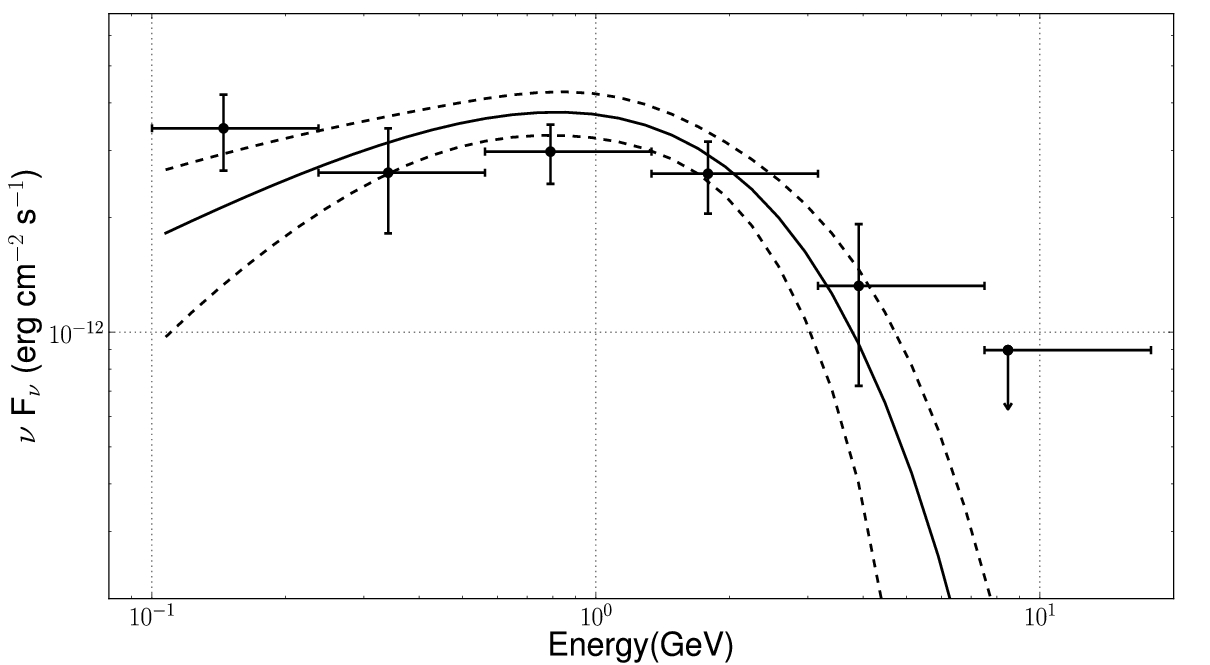}
  \end{center} {\bf Fig.~S1.} $\gamma$-ray spectral energy distribution of \psr\
obtained with the Fermi Large Area Telescope. The solid black line shows the
maximum likelihood fit to a power law with exponential cut-off. The
dashed lines are $\pm1 \sigma$ uncertainties on the fit parameters. Plotted points
are from likelihood fits to individual energy bands with  $>3 \sigma$ detection
above background for two degrees of freedom, otherwise a 95\% confidence
level upper limit arrow is shown. The errors are statistical only.
\end{figure}

\begin{figure}[htbp]
  \begin{center}
     \includegraphics[width=16cm]{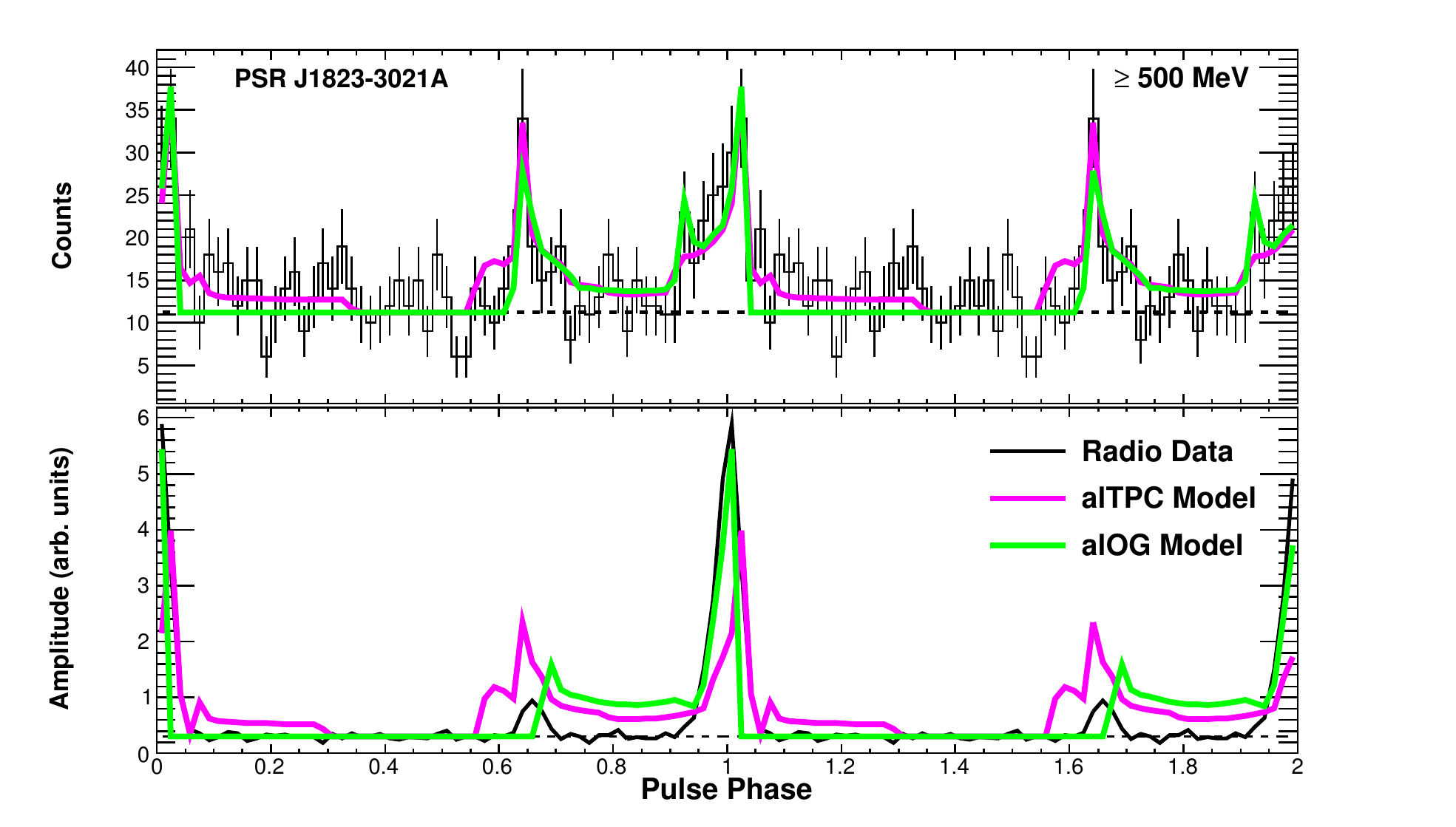}
  \end{center} {\bf Fig.~S2.} Observed and best-fit $\gamma$-ray (top) and radio (bottom) light 
curves for PSR~J1823$-$3021A using the alTPC (pink) and alOG (green) models described 
in the text. The dashed, horizontal lines in both panels correspond to the estimated background 
levels. The $\gamma$-ray background was estimated using an annular ring centered on the radio 
position with inner and outer radii of 1$^{\circ}$ and 2$^{\circ}$, respectively. The radio 
background was estimated by fitting the region between 0.1 and 0.6 in phase to a constant 
value. The parameters of the best-fit light curves are given in Table S2.
\end{figure}


\end{document}